\documentclass[letter]{aa}

\usepackage{graphicx}
\usepackage{subcaption}
\usepackage{amsmath}
\usepackage{textcomp, gensymb}
\usepackage{tabularx}
\usepackage{booktabs}
\usepackage{lipsum}
\usepackage{comment}
\usepackage{txfonts}
\usepackage{hyperref}
\hypersetup{
    colorlinks=true,
    linkcolor=blue,
    filecolor=magenta,      
    urlcolor=blue,
    citecolor=blue
}
\usepackage{enumitem}
\usepackage[normalem]{ulem}
\usepackage{changepage}
\usepackage[para]{threeparttable}
\usepackage[switch]{lineno}

\usepackage{array}
\usepackage{changepage}
\usepackage{makecell}
\usepackage{orcidlink}
\usepackage{etoolbox}

\RequirePackage{soul}
\RequirePackage{xcolor} 

\newcommand{\target}{5BZQ~J1243$+$4043}

\newcommand{\MgII}{Mg~\MakeUppercase{\romannumeral2}}
\newcommand{\CIV}{C~\MakeUppercase{\romannumeral4}}
\newcommand{\CIII}{C~\MakeUppercase{\romannumeral3}]}

\newcommand{\NIII}{N~\MakeUppercase{\romannumeral3}]}
\newcommand{\HeII}{He~\MakeUppercase{\romannumeral2}}

\newcommand{\FeI}{Fe~\MakeUppercase{\romannumeral1}}
\newcommand{\accretion}{L_{\rm BLR}/L_{\rm Edd}}

\newcommand{\Lblr}{L_{\rm BLR}}

\newcommand{\redshift}{$z = 1.5181\pm0.0002$}
\newcommand{\Pradio}{P_{1.4\,{\rm GHz}}}
\newcommand{\Funits}{${\rm erg}\cdot{\rm cm}^{-2}\cdot{\rm s}^{-1}\cdot$\AA$^{-1}$}

\newcommand{\WHz}{{\rm W}\cdot{\rm Hz}^{-1}}

\newcommand{\origin}{\textsc{Paper~I}}
\newcommand{\originII}{\textsc{Paper~II}}

\newcommand{\paperPHone}{\textsc{Paper~PI}}

\makeatletter
\renewcommand*\aa@pageof{, page \thepage{} of \pageref*{LastPage}}
\makeatother

\begin{document}

\title{The changing look of the neutrino-emitter blazar candidate \target}

   \subtitle{}

   \author{Alessandra Azzollini\orcidlink{0000-0002-2515-1353}
          \inst{1, 2}
          \and
            Sara Buson\orcidlink{0000-0002-3308-324X}
          \inst{1,2}
          \and 
            Alexis Coleiro\orcidlink{0000-0003-0860-440X}
          \inst{3}
          }

   \institute{Julius-Maximilians-Universit\"at W\"urzburg, Fakultät f\"ur Physik und Astronomie, Institut f\"ur Theoretische Physik und Astrophysik, Lehrstuhl f\"ur Astronomie, Emil-Fischer-Str. 31, D-97074 W\"urzburg, Germany\\
        \email{alessandra.azzollini@uni-wuerzburg.de}
        \and 
        Deutsches Elektronen-Synchrotron DESY, Platanenallee 6, 15738 Zeuthen, Germany
        \and
        Université Paris Cité, CNRS, Astroparticule et Cosmologie, F-75013 Paris, France}

   \date{Received 21/09/2025; accepted 28/10/2025}

  \abstract
   {In recent years, changing-look blazars have called the traditional view of BL Lacs-flat spectrum radio quasars into question within the empirical classification of blazars. Based on the intensity of the optical spectral lines, they appear to transition between the two classes over time.}
   {We focus on the blazar \target, recently proposed as a promising candidate for the emission of high-energy neutrinos observed by the IceCube Neutrino Observatory and reported as a changing-look blazar in the literature. We study the spectral properties of this blazar, inferring its radiation field and accretion regime properties across different epochs.}
   {This study presents new optical spectroscopy observations of \target\ taken with the Gran Telescopio Canarias. We used this new dataset and two optical spectra available from the literature to investigate the continuum and line emissions and pinpoint the physical properties of the source. In particular, we used the emission lines to probe the accretion regime.} 
   {The newly collected data for \target\ shows broad emission lines, 
    consistent with the spectrum of the first epoch and the redshift \redshift\ known from the literature. For the second epoch, the spectrum appears featureless and so, we placed limits on the emission lines and related physical properties. 
    We observed spectral variability for both the continuum and line emissions among the three spectra. Nonetheless, the accretion properties of the blazar generally remain unvaried, indicating that the intrinsic physics stays the same across the three epochs. In the broader multi-messenger context, this suggests that, despite the changing look in the optical band, the candidate neutrino-emitter blazar \target\ is still characterized by the presence of intense external radiation fields and radiatively efficient accretion, typical of high-excitation radio galaxies, which may foster neutrino production.}
   {}


   \maketitle
%

\section{Introduction}
\label{sec: intro}

Active galactic nuclei (AGNs) are among the most luminous, persistent extragalactic sources in the Universe and are powered by accretion onto the central supermassive black hole (SMBH). Among them, the radio-loud subclass of blazars hosts highly relativistic jets pointed in the observer's line of sight. 
Blazars may exhibit optical spectral lines produced when photons emitted by the accretion disk interact with surrounding gas clouds. These include the fast-moving clouds of the broad-line region (BLR), located closer to the central black hole, and the slower-moving gas of the narrow-line region (NLR) farther out. The higher-velocity denser \citep[$n_{\rm e}\gtrsim10^9\,{\rm cm}^{-3}$; ][]{Osterbrock} gas of the BLR is responsible for broad lines (with a   full width at half maximum, FWHM, of $\gtrsim1000\,{\rm km}\cdot{\rm s}^{-1}$) of permitted and semi-forbidden transitions.
The slower and less dense NLR \citep[$n_{\rm e}\lesssim10^{6}\,{\rm cm}^{-3}$; ][]{Longair:book}  give origin to narrow forbidden lines instead.

Within the AGN unification model \citep{Urry:1995, Urry:2004}, blazars are traditionally classified in BL Lacertae objects (BL Lacs) and flat spectrum radio quasars (FSRQs) based on the rest-frame equivalent width (EW) of the optical spectral lines, which measures the line strength compared to the underlying continuum. In this observational-based scheme, FSRQs display intense broad (EW $> 5$~\AA) emission lines in their optical spectra,  while BL Lacs lack emission lines or show only absorption and narrow emission lines (EW $< 5$~\AA). 

A more physically driven scenario based on the modes of accretion onto the central SMBH and the intrinsic power of the relativistic jet has been proposed in the literature \citep{Ghisellini_Celotti:2001, BestHeckman:2012, Giommi:2012}. This approach was also adopted in our first work on the physical properties of candidate neutrino-emitter blazars \citep[][hereafter, \paperPHone]{Azzollini_2025}. The accretion efficiency is traced by the ratio of the BLR luminosity in Eddington units, $\accretion$ \citep{Ghisellini:2009, Ghisellini:2011, Sbarrato:2012, Sbarrato:2014}. The radio power at $1.4\,{\rm GHz}$, instead, traces the intrinsic jet's power \citep{BestHeckman:2012, Heckman:2014, Padovani:2022}.
Based on their intrinsic properties, high-excitation radio galaxies (HERGs) are characterized by intense external radiation fields, efficient accretion, and powerful jets that exceed these thresholds. In contrast, low-excitation radio galaxies (LERGs) show weak or absent optical emission lines, reflecting their inefficient accretion, weaker radiation fields, and reduced jet power. The proposed dividing thresholds are $\accretion\sim5\times10^{-4}$ \citep{Ghisellini:2010, Ghisellini2011_highred} and $\Pradio\sim10^{26}\,\WHz$ \citep{BestHeckman:2012}. 
Following this physical classification, FSRQs and BL Lacs could be considered as the jetted counterparts of HERGs and LERGs, respectively.

In recent years, the empirical BL Lacs/FSRQs scenario has been called into question by blazars that transition between the two classes over time. These objects, referred to as changing-look blazars, exhibit emission lines in their optical spectra during some epochs, while appearing featureless at others \citep[][]{Vermeulen:1995, Ghisellini:2011, Ruan:2014, PenaHerazo_changinglook,Ricci_2023}. Several hypotheses have been proposed to explain this phenomenon. 
It could be associated with sudden changes in the accretion rate: broad spectral profiles appear (disappear) concurrently with the sudden increase (decrease) in the accretion rate \citep{Xiao:2022, Dong_2024, Duffy_2025}.
Alternatively, Doppler boosting may modify the intensity and beaming of the photo-ionizing continuum over time and, consequently, this affects the interplay with the thermal line emission. 
Some studies have shown that the EW of the BLR lines is anti-correlated with the continuum luminosity \citep{Corbett:2000, Ruan:2014, Paiano:2024}. In this framework, the variable jet-driven non-thermal continuum can outshine the BLR emission, making the broad lines temporarily undetectable.

In \paperPHone, we investigated the nature of the $52$ blazars proposed as promising candidates of neutrino emission in \citet[][hereafter, \origin]{Buson:2022, Buson_erratum} and \citet[][hereafter, \originII]{Buson:2023}. Investigations of their HERG/LERG properties have revealed that the sample includes representatives of both classes, with a mild tendency toward HERG-like properties. In the lepto-hadronic scenario, the efficiency of photo-pion production, which is a key mechanism for neutrino production, is strongly influenced by the accretion mode. Identifying the presence of external radiation fields, such as those characteristic of HERGs, is therefore important when we are looking to interpret the physical conditions leading to neutrino emission.

The blazar \target\ was part of that sample, it was proposed as the blazar counterpart of a $\gamma$-ray source in \cite{Principe_2021} and was classified as a HERG based on optical spectra from 2012. 
However, prior studies have suggested it may be a changing-look blazar, exhibiting transitions between the FSRQ and BL Lac classes due to variability in its optical emission lines over time \citep{PenaHerazo:2021}. This prompts the question of whether the observed spectral changes reflect a genuine evolution in the blazar’s physical properties, requiring a reassessment of its HERG$-$LERG classification with implications for neutrino production or cosidering whether they might instead be the result of observational effects with limited physical implications.

In this paper, we focus on the optical spectroscopic properties of the blazar \target, using both archival and new optical spectra among three epochs.
The work is organized as follows. Section \ref{sec: data analysis} describes the dataset and analysis. Section \ref{sec: discussion} presents the findings. Section \ref{sec: conclusions} summarizes our conclusions. We assumed a flat $\Lambda$CDM cosmology with $\rm{H}_{\rm 0} = 69.3~\rm{km}\cdot{\rm s}^{-1}\cdot\rm{Mpc}^{-1}$, $\Omega_{\rm m0} = 0.29$, and $\Omega_{\Lambda} = 0.71$.

\section{Optical spectroscopy analysis}
\label{sec: data analysis}

\subsection{Optical spectroscopy dataset}
\label{subsec: dataset}

\begin{figure}[t]
    \centering
    \includegraphics[width = 0.8\columnwidth]{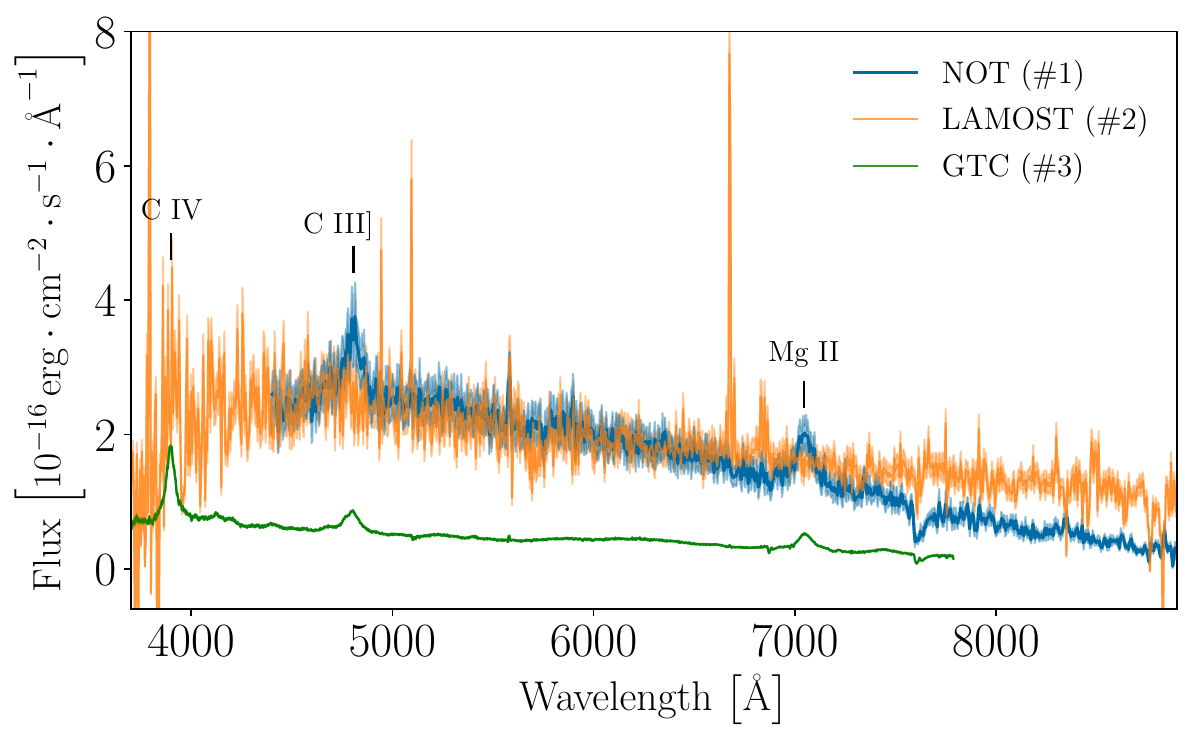}
    \caption{\footnotesize Optical spectra of \target\ from: the NOT spectrum \#1 \citep[MJD $56035.11$; ][]{Titov:2013} shown in blue, the LAMOST spectrum \#2 \citep[MJD $56360.75$; ][]{PenaHerazo_changinglook} in orange, and a new GTC spectrum \#3 acquired in this work (MJD $60318.21$) in green. The shaded areas indicate the corresponding flux uncertainties.}
    \label{fig: spectra}
\end{figure}

Three optical spectra are available for \target\ at different times spanning $\sim$12 years (shown in Fig. \ref{fig: spectra}).
A first spectrum was acquired with the Nordic Optical Telescope (NOT, blue)
on April $18$th, $2012$ \citep[MJD $56035.11$, hereafter spectrum \#1;][]{Titov:2013}.

Based on the detection of prominent emission lines in spectrum \#1, the blazar \target\ was classified as FSRQ in the fifth edition of the Roma-BZCat catalog \citep[$5$BZCat;][]{BZCat}, with a redshift of \redshift.
This spectrum was used in \paperPHone\ to determine the physical classification of \target, identifying it as an HERG.

A second spectrum was acquired on March $9$th, $2013$ \citep[MJD $56360.75$, hereafter spectrum \#2;][]{PenaHerazo:2021} within the Large Sky Area Multi-Object Fiber Spectroscopic Telescope  Survey \citep[LAMOST;][]{LAMOST:2012}. This spectrum was pointed out as featureless and our analysis confirmed these earlier findings, based on the spectral components being narrower than the minimum width set by the instrument resolution.

We acquired a third spectrum on January $9$th, $2024$ (MJD $60318.21$; hereafter, spectrum \#3) with the R$1000$B grism of the Optical System for Imaging and low-Intermediate-Resolution Integrated Spectroscopy of the Gran Telescopio Canarias \citep[GTC OSIRIS, green;][]{osiris}. We reduced this spectrum using the \texttt{PypeIt v1.16.0} pipeline \citep{Pypeit} following standard analysis procedures 

\subsection{Methodology for the estimation of physical properties}
\label{subsec: properties}

The three spectra were examined and used for further analysis following the same approach described in \paperPHone.
After extracting the three spectra, we applied a power-law fit $F_{\lambda}\propto \lambda^{\alpha}$ to the spectral continuum around the \MgII\ line profile \citep[as in][]{Pandey_2025} to model the non-thermal radiation of the jet \citep{Falomo:1994, Urry:1995, Scarpa:1997, Corbin:1997, Ghisellini:1998}. We analyzed the identified emission lines and extracted the flux, EW, and FWHM.
For the featureless spectrum \#2, we estimated upper limits (ULs) on the undetected lines.
The total luminosity of the emission lines (or the limit) was used to infer the accretion BLR luminosity $\Lblr$ of the blazar.
Furthermore, we used the lines' luminosity and FWHM to estimate the virial black hole mass and the Eddington luminosity as a result \citep[as in \paperPHone; see also][]{Shen:2011, McLure:2004, Vestergaard:2009}. As in \paperPHone, we used the radio flux density at $1.4\,{\rm GHz}$ from the NRAO VLA Sky Survey \citep[NVSS,][]{NVSS} to determine the radio power that traces the intrinsic power of the relativistic jet.

\subsection{Results}
\label{subsec:results}

\begin{figure}
    \centering
    \includegraphics[width = 0.7\columnwidth]{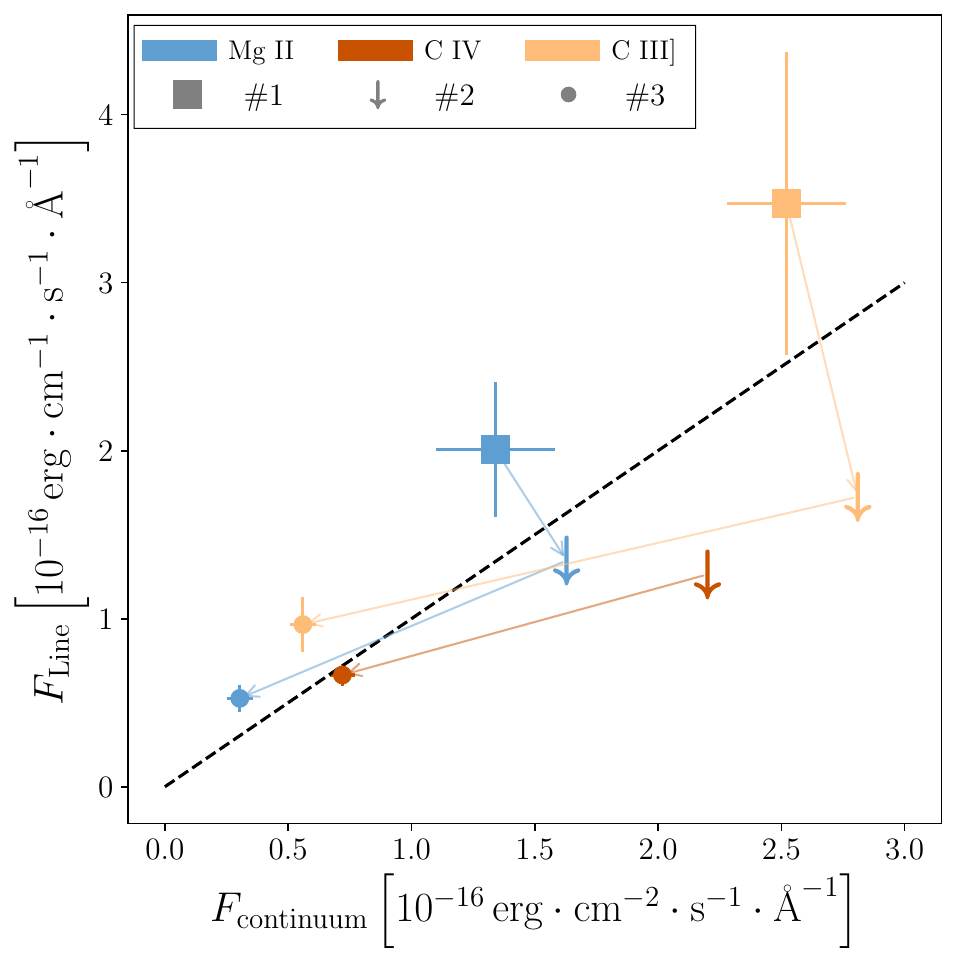}
    \caption{\footnotesize Variations in the emission line fluxes as a function of the underlying continuum level across the three epochs of \target. The values for \MgII, \CIV, and \CIII\ emission lines are indicated in blue, brown, and yellow, respectively. The values corresponding to  spectra \#1 and \#3 are indicated with a square and a filled circle, respectively, while the arrows show the limits derived for epoch \#2. The dotted black line traces the diagonal, corresponding to equality.}
    \label{fig: lines vs. continuum}
\end{figure}

The continuum and line emission estimates are listed in Table \ref{table: properties} and displayed in Fig. \ref{fig: lines vs. continuum} (see also Fig. \ref{fig: continuum fit}). 
In the NOT spectrum \#1, we confirmed the detection of broad \CIII\ and \MgII\ emission lines at $4796$ and $7040$ \AA, respectively, consistent with the redshift \redshift\ \citep{Titov:2013}. The continuum is described by a power law with a flux of$\left(1.34\pm0.24\right)\times10^{-16}\,{\rm erg}\cdot{\rm cm}^{-2}\cdot{\rm s}^{-1}\cdot{\rm \AA}^{-1}$ and spectral index of $\alpha = -1.51\pm0.12$. 
Spectrum \#2 appears featureless in terms of emission lines, consistent with the object exhibiting BL Lac characteristics at that time, thereby leading to its classification as a changing-look blazar. We identified two absorption components at $\sim8766$ and $\sim8833$ \AA, likely corresponding to \FeI\ transitions \citep{Meggers_1961}. This remains consistent with the non-detection of BLR lines. The continuum follows a power law with a flux of $\left(2.21\pm0.49\right)\times10^{-16}\,{\rm erg}\cdot{\rm cm}^{-2}\cdot{\rm s}^{-1}\cdot{\rm \AA}^{-1}$ and index of $\alpha = -1.00\pm0.24$. We placed an UL on the BLR line fluxes by simulating \CIV, \CIII, and \MgII\ lines. 

In the newly acquired spectrum \#3, we identified \CIV\ at $3898$ \AA, \HeII\ $1640$ \AA\ at $4128$ \AA, \NIII\ $1750$ \AA\ at $4405$ \AA, \CIII\ $1909$ \AA\ at $4802$ \AA,\, and \MgII\ $2798$ \AA\ at  $7046$ \AA. These results confirm the redshift value \redshift. The \MgII\ profile shows two double absorption components on its blue wing, at $6871-6890$ \AA\ and $6975-6993$ \AA, respectively, which might be associated with the interstellar medium (ISM) of the host galaxy, with intervening absorbing material from massive galaxies nearby and/or the presence of wind outflowing from the disk \citep{Santoro:2018, Santoro:2020, Paiano:2024, Chen:2025}. The continuum follows a power-law of flux $\left(0.30\pm0.05\right)\times10^{-16}\,{\rm erg}\cdot{\rm cm}^{-2}\cdot{\rm s}^{-1}\cdot{\rm \AA}^{-1}$ and index $\alpha = -1.50\pm0.08$.

\begin{table}[t]
\centering
\caption{Spectral properties of \target\ across the three epochs.}
\resizebox{\linewidth}{!}{
\begin{tabular}{llccc}
\toprule
Spectrum & Component & \thead{Flux \\ $10^{-16}\,$\Funits} & \thead{rest-frame EW \\ \AA} &  \thead{FWHM \\ ${\rm km}\cdot{\rm s}^{-1}$}\\
\hline
NOT \#1     &   Continuum       &   $1.34\pm 0.24$    &   $-$       &   $-$         \\
        &   \CIII\           &   $3.47\pm 0.90$    &   $23.43\pm 2.11$   &   $6007.87\pm 90.12$   \\
        &   \MgII\           &   $2.01\pm 0.40$    &   $24.87\pm 2.98$   &   $6972.74\pm 146.43$   \\
\hline
LAMOST \#2     &   Continuum       &   $2.21\pm 0.49$    &   $-$       &   $-$         \\
        &   \CIV\            &   $<1.26$   &  $-$   &   $4000.00^a$   \\ 
        &   \CIII\           &   $<1.72$   &  $-$   &   $4000.00^a$   \\ 
        &   \MgII\           &   $<1.35$   &  $-$   &   $4000.00^a$   \\ 
\hline
GTC \#3     &   Continuum       &   $0.30\pm 0.05$    &   $-$       &   $-$         \\
        &   \CIV\            &   $0.66\pm 0.07$    &   $36.30\pm 4.36$   &   $4774.69\pm 42.97$   \\
        &   \HeII           &   $0.25\pm 0.04$    &   $1.74\pm 0.14$    &   $1254.48\pm 3.76$   \\
        &   \NIII           &   $0.77\pm 0.13$    &   $2.82\pm 0.23$    &   $1499.47\pm 7.50$   \\
        &   \CIII\           &   $0.96\pm 0.14$    &   $25.10\pm 3.26$   &   $6152.56\pm 86.14$   \\
        &   \MgII\           &   $0.53\pm 0.08$    &   $51.87\pm 10.37$   &   $5628.52\pm 157.60$ \\
\hline
\end{tabular}}
\tablefoot{The first column reports the reference spectrum.
The second, third, fourth and fifth columns list the spectral component with corresponding flux, either measured or limits, rest-frame EW and FWHM. \\
$^a$ In spectrum \#2, we placed ULs on the undetected emission lines by fixing the FWHM (see \paperPHone).
}
\label{table: properties}
\end{table}

\section{Discussion}
\label{sec: discussion}

The three observations of \target\ used in this study point toward a variation in the BLR luminosity over $\sim12$ yr, with a gradual decrease of the line luminosities while the accretion regime remains, overall, unvaried. We observed variations in both the continuum and line emission, as evident from Fig.~\ref{fig: lines vs. continuum}. The continuum flux increases from spectrum \#1 to spectrum \#2, when it reaches its maximum coincident with a featureless optical spectrum; then it decreases again in spectrum \#3. A first implication of the observed spectral variability is that this blazar is not characterized by a straightforward disk-jet relation: when BLR lines are detected, in epoch \#1 and \#3, both the line and the continuum flux drop, while the corresponding EW increases. This suggests that the continuum (i.e., the jet emission) decreases more rapidly than the line (disk) emission. 
During the higher jet state, \#2, the question of whether the BLR lines are outshone or their absence reflects a drop in the disk luminosity (and, hence, a change in the radiation field properties) remains open. To study this, we performed a test simulation to investigate the impact of spectral lines as bright as those observed in \#1 and \#3 over a continuum as high as spectrum \#2; namely, the highest observed for this blazar and with no evidence of lines. As shown in Figs. \ref{fig: simulation Mg II} and \ref{fig: simulation C III}, we found that the presence of BLR lines as luminous and broad as those detected in $\#1$ and $\#3$ is not excluded in spectrum $\#2$. If present, they would be outshone by the optical continuum flux, likely dominated by the non-thermal component.
An overall higher jet activity contemporaneous with epochs \#1 and \#2 is consistent with the $\gamma$-ray detection of the blazar during 2008–2019 \cite{Principe_2021} and its non-inclusion in more recent $Fermi$-LAT catalogs \citep{4FGL-DR4}.

\begin{figure}
    \centering
    \includegraphics[width = 0.8\columnwidth]{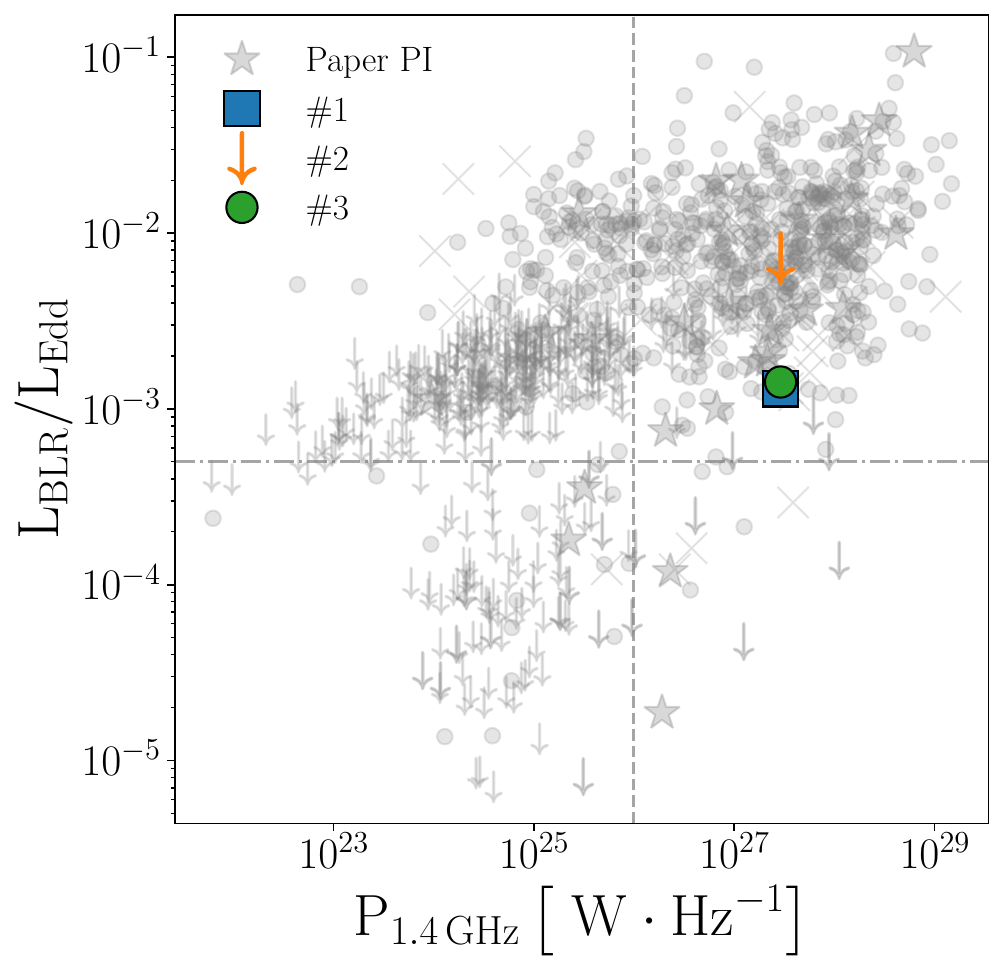}
    \caption{\footnotesize Accretion regime $\accretion$ as a function of the radio power at $1.4$ GHz. The three considered epochs (\#1 as a blue square, \#2 as an orange arrow, \#3 as a green dot) of \target\ are compared to the blazar samples analyzed in \paperPHone\ (gray). The horizontal dotted and vertical dashed-dotted black lines represent the boundaries for the LERG/HERG-like behavior, respectively. As discussed in Sect. \ref{sec: discussion}, despite the line emission variations, the object remains in the radiative-efficient regime.}
    \label{fig: accretion regime}
\end{figure}

Despite the observed variability in the line and continuum emission across the three epochs, the accretion properties and, more generally, the underlying physics, ultimately remain the same. This is traced by the $\accretion$ ratio, which \ lies above the dividing line for \target\  \citep[$\accretion\sim5\times10^{-4}$;][]{Ghisellini:2011, BestHeckman:2012, Sbarrato:2014}. In other words, it displays HERG-like properties: the ratio is $\accretion\sim 1.30\times10^{-3}$ in \#1 and $\sim 1.43\times10^{-3}$ in \#3. In epoch \#2, the ULs result in $\accretion\lesssim 7.19\times10^{-3}$. Similarly, the jet power of this blazar is $\Pradio \sim 2.91\times10^{27}\WHz$, above the corresponding threshold $\Pradio\simeq 10^{26}\,\WHz$ \citep{BestHeckman:2012, Padovani:2022}, which corroborates its HERG-like nature. 
This is also shown in Fig. \ref{fig: accretion regime}, which represents the accretion regime $\accretion$ as a function of $\Pradio$. The values of three epochs (\#1 as a blue square, \#2 as an orange arrow, and \#3 as a green dot) of \target\ are overlaid to those of the candidate neutrino-emitter blazar sample and other reference blazar samples investigated in \paperPHone\ (gray). Horizontal dashed and vertical dashed-dotted lines indicate the LERG$-$HERG thresholds.

This shows that despite the observed spectral variability, the nature of the candidate neutrino-emitter blazar \target\ remains consistent with radiatively efficient accretion, intense external radiation fields, and high jet power across the $\sim12$ yr. This is consistent with the mild tendency toward HERG-like properties found for the full sample of neutrino candidates (\paperPHone; see also \origin\ and \originII).

\section{Summary and conclusions}
\label{sec: conclusions}

In this work, we studied the behavior of the blazar \target, which was recently proposed as a candidate for IceCube neutrinos and as a changing-look blazar in the literature, among three epochs covered by optical spectroscopy observations. We used new data acquired with the Gran Telescopio Canarias for the third epoch. We analyzed the properties of the emission lines detected in the first and third spectra and placed limits on the nondetections during the second epoch. We compared their flux with the continuum in all three cases.

Our analysis shows that the highest continuum coincides with a featureless spectrum. However, this is not inconsistent with the presence of BLR lines, as observed for the other two epochs.
We observed spectral variability in both the continuum and emission lines, with the decrease in the line intensities between the first and third epoch accompanied by an increase in the corresponding equivalent width and a slight hardening of the spectral index. 

Our findings show that while observational properties vary across the three epochs, the underlying physical properties remain the same. Among the three considered epochs, the candidate neutrino-emitter blazar \target\ is characterized by intense external radiation fields, a radiatively efficient accretion regime, and powerful jets that are typical of HERG-like sources. This is particularly important in the broader multi-messenger context, since \target\ has been pointed out as a promising candidate for the long-term emission of IceCube neutrinos (\originII). The combination of efficient particle acceleration and the presence of intense external radiation fields might indeed foster neutrino production in the powerful jet of efficiently accreting blazars, despite the observed variations in the appearance of spectral lines in the optical band.

\begin{acknowledgements}
{\scriptsize We thank the anonymous referee for the constructive feedback. We thank Jose Maria Sanchez Zaballa and Leonard Pfeiffer for the insightful comments and fruitful discussions. This work was supported by the European Research Council, ERC Starting grant \emph{MessMapp}, S.B. Principal Investigator, under contract no. $949555$.}
{\scriptsize This work is (partly) based on data obtained with the instrument OSIRIS, built by a Consortium led by the Instituto de Astrofísica de Canarias in collaboration with the Instituto de Astronomía of the Universidad Autónoma de México. OSIRIS was funded by GRANTECAN and the National Plan of Astronomy and Astrophysics of the Spanish Government.
}
\end{acknowledgements}
\vspace{-0.7cm}

\bibliographystyle{aa}
\bibliography{biblio}

\newpage

\begin{appendix}

\section{Test simulation of the emission lines}
\label{sec: appendix line}

\begin{figure}[h]
    \centering
    \includegraphics[width = \columnwidth]{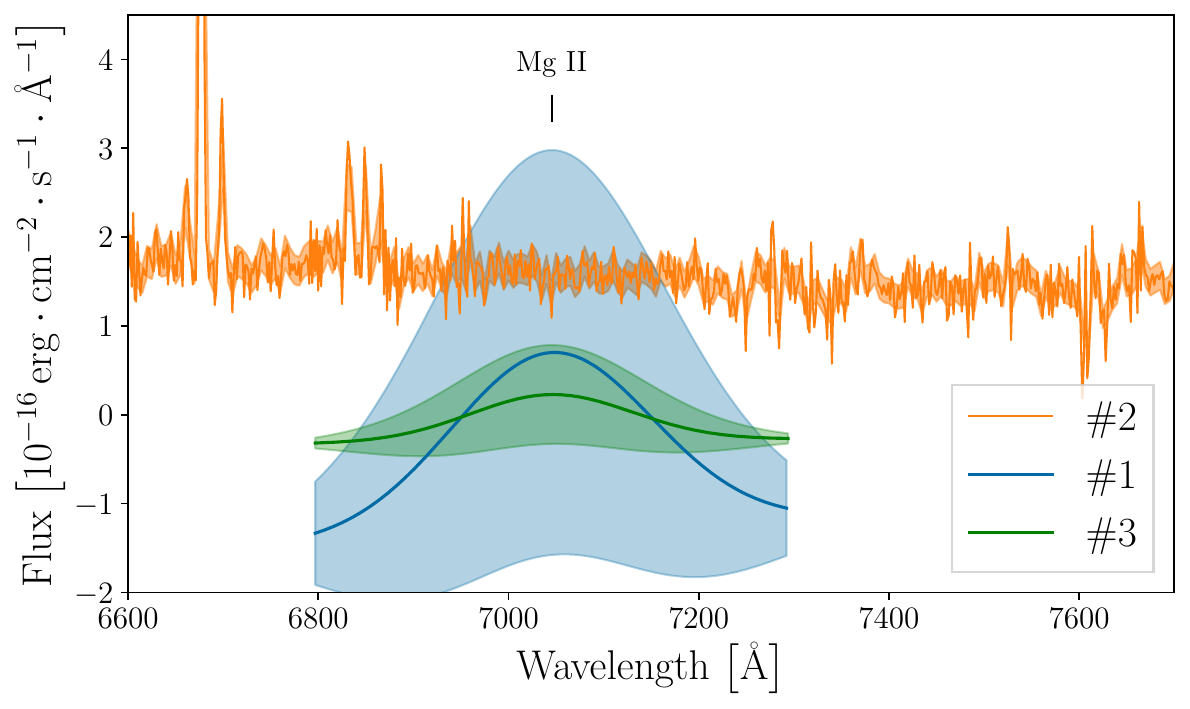}
    \caption{Test simulation of a \MgII\ as bright and broad as in spectra \#1 (blue) and \#3 (green) over a continuum as high as in \#2 (orange). The shaded areas show the corresponding $1\sigma$ uncertainties.}
    \label{fig: simulation Mg II}
\end{figure}

\begin{figure}[h]
    \centering
    \includegraphics[width = \columnwidth]{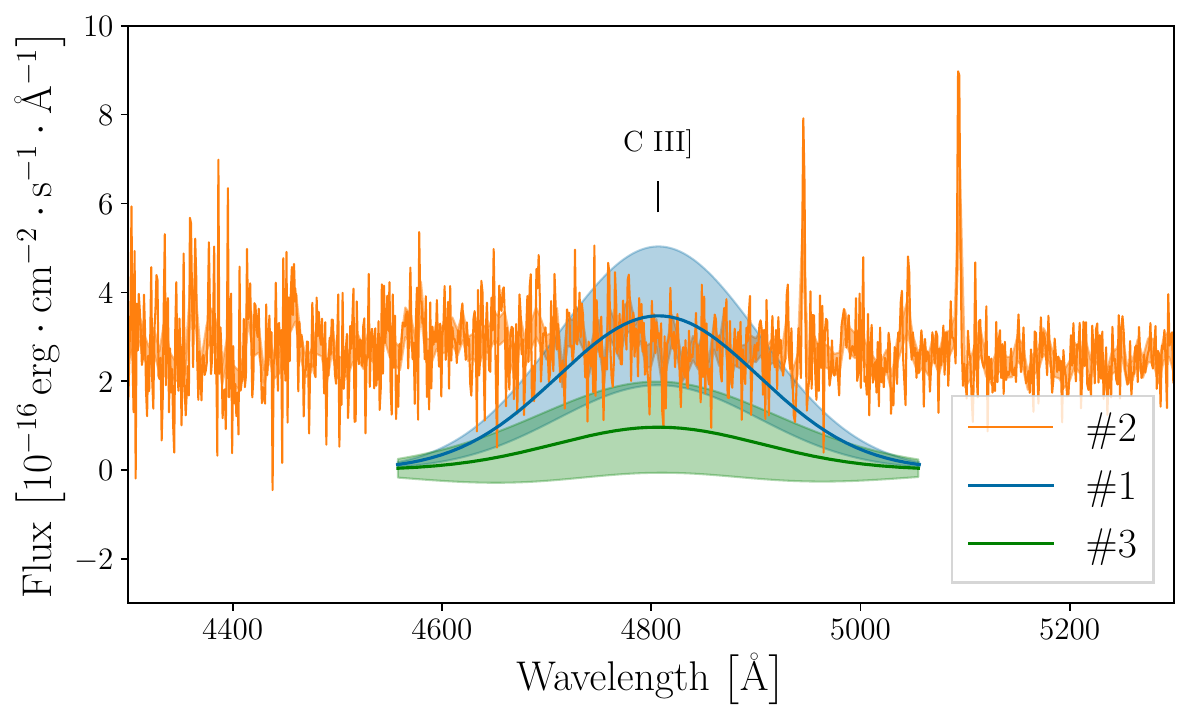}
    \caption{Test simulation of a \CIII\ as bright and broad as in spectra \#1 (blue) and \#3 (green) over a continuum as high as in \#2 (orange). The shaded areas show the corresponding $1\sigma$ uncertainties.}
    \label{fig: simulation C III}
\end{figure}

We aim to investigate the line variability across the three epochs of \target. To this end, we simulated a Gaussian profile corresponding to a \MgII\ and \CIII\ emission lines with the same flux, EW and FWHM as the ones observed in spectra \#1 and \#3. Then, we compared their brightness to a continuum as high and hard as in spectrum \#2, i.e., the highest observed for this blazar and with no evidence of lines. The result is shown in Figs. \ref{fig: simulation Mg II}-\ref{fig: simulation C III}, which shows the spectrum $\#2$ in orange, and the Gaussians corresponding to the \MgII\ and \CIII\ fluxes of $\#1$/$\#3$ in blue/green, respectively. The shaded areas indicate the corresponding $1\sigma$ uncertainties on the flux.

This result shows that if lines as bright as those seen in spectra $\#1$/$\#3$ are present, they would be outshone by the higher jet continuum.

\section{The fit of the continuum}
\label{sec: pwl fit}

\begin{figure}[h]
    \centering
    \includegraphics[width = \columnwidth]{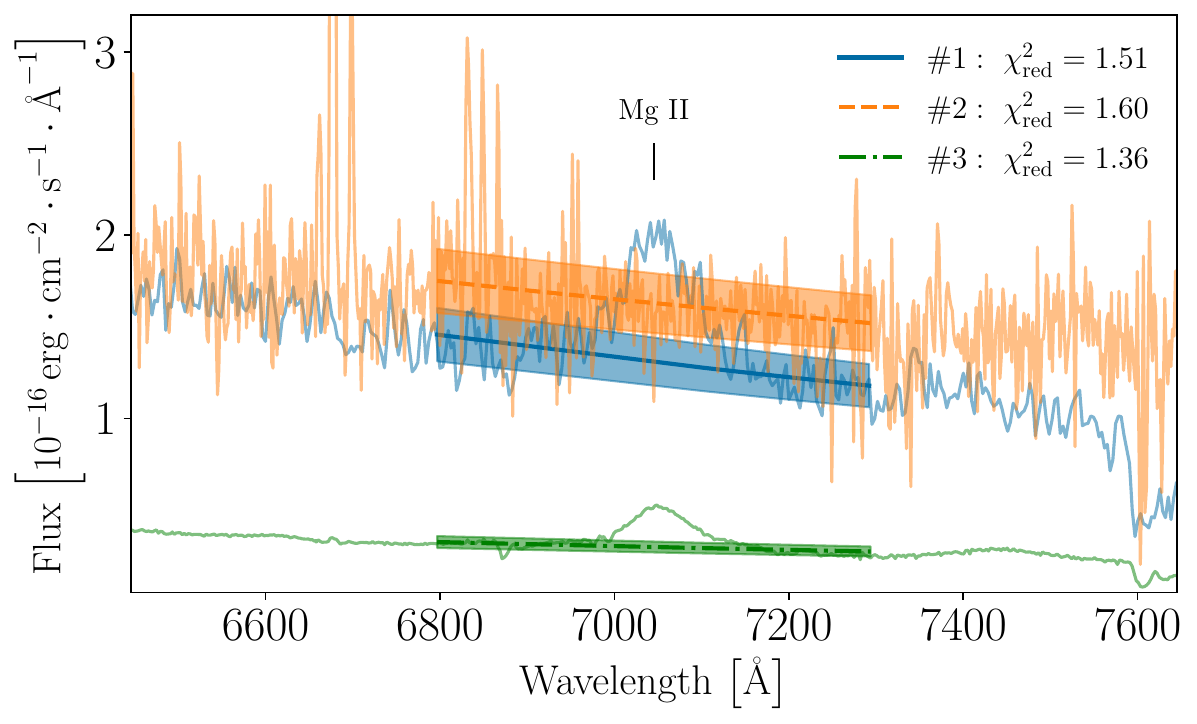}
    \caption{Power-law fit of the spectral continuum in the three epochs of \target\ with the corresponding reduced $\chi^2_{\rm red}$: \#1 shown in solid blue, \#2 in dotted orange, and \#3 in dash-dotted green. The shaded areas indicate the corresponding flux uncertainties.}
    \label{fig: continuum fit}
\end{figure}

As explained in Sect. \ref{subsec: properties}, we applied a power-law fit to the spectral continuum around the \MgII\ line profile to investigate the changes in the non-thermal radiation of the jet across the three epochs. The results are reported in Table \ref{table: properties}, and shown in Fig. \ref{fig: continuum fit}. 

\section{Journal of observations}
\label{sec: journal of observations}

Table \ref{tab: journal of observations} provides further details on the spectral dataset used for the analysis. The three observations were performed under the same conditions of slit width greater than or similar to the seeing, ensuring a reliable flux calibration and estimation\footnote{For further details on GTC OSIRIS see, for example, \url{https://atmosportal.gtc.iac.es} and \url{https://www.gtc.iac.es/instruments/osiris+/osiris+.php}.} \citep{Titov:2013, Song_2012_seeing}.

\begin{table}[ht]
\centering
\caption{Observational details of the spectral dataset.}
\resizebox{\linewidth}{!}{
\begin{tabular}{llccc}
\toprule
Spectrum & Instrument & Resolution & Observation date [MJD] & $\mathbf{t_{\rm exp}}\left[{\rm ks}\right]$\\
\hline
\#1 & NOT ALFOSC$^1$ & $415$ & $56035.11$ & $2.40$ \\
\#2 & LAMOST$^2$ & $1000$ & $56360.75$ & $6.37$ \\
\#3 & GTC OSIRIS & $1018$ & $60318.21$ & $3.00$ \\
\hline
\end{tabular}}
\tablefoot{The first column reports the reference spectrum. The second, third, fourth and fifth columns list the observation details with corresponding instrument, resolution $R=\lambda/\Delta \lambda$, observation date and total exposure time. \\
\tablefoottext{1}{From \cite{Titov:2013}}
\tablefoottext{2}{From \cite{PenaHerazo_changinglook}}
}
\label{tab: journal of observations}
\end{table}

\end{appendix}

\end{document}